\documentstyle{article}
\begin{document}
\newcommand{\eqb}{\begin{equation}}
\begin{center}
{\Large{\bf The myth of the photon}}\vspace{1 cm}\\
Trevor W. Marshall, Dept. of Mathematics, Univ. of Manchester,\\
Manchester M13 9PL, UK\\
Emilio Santos, Depto de F\'isica Moderna, Univ. de Cantabria,\\
39005 Santander, Spain
\end{center}
{\bf Abstract}

We have shown that all ``single-photon'' and ``photon-pair''
states, produced in atomic transitions, and in parametric
down conversion by nonlinear optical crystals,
may be represented by positive Wigner
densities of the relevant sets of mode
amplitudes. The light fields of all such states are represented
as a real probability ensemble (not a pseudoensemble) of
solutions of the unquantized Maxwell equation.

The local realist analysis of
light-detection events in spatially separated
detectors requires a theory of detection which goes beyond the
currently fashionable single-mode ``photon'' theory. It also
requires us to recognize that there is a payoff between
detector efficiency and signal-noise discrimination. Using
such a theory, we have demonstrated that all experimental
data, both in atomic cascades and in parametric down conversions,
have a consistent local realist explanation based on the
unquantized Maxwell field.

Finally we discuss current attempts to demonstrate
Schroedinger-cat-like behaviour of microwave cavities interacting with
Rydberg atoms. Here also we demonstrate that there is no
experimental evidence which cannot be described by the
unquantized Maxwell field.

We conclude that the ``photon'' is an obsolete concept,
and that its misuse has resulted in a mistaken
recognition of ``nonlocal'' phenomena.

(This article was published in "The Present Status of the
Quantum Theory of Light", ed. S. Jeffers et al., Kluwer,
Dordrecht, 1997, pages 67 - 77)

\section{Introduction}

Since the beginning of modern physics, that is since the
seventeenth century, there have been two views as to the
nature of light. The corpuscular view has, traditionally,
been supported by those most strongly attracted to formal
elegance, whilst the undulatory view has been supported by those who
insist on the necessity for science to give explanations
of phenomena $^{(1)}$. It is no coincidence that  one of the
earliest and strongest statements against trying to explain
field phenomena is in Newton's {\it Opticks}, and that a similar
statement was made, nine years later, in the Preface to his
{\it Principia}.

Formal elegance, combined with Newton's authority,
dominated the eighteenth and the first part of the
nineteenth century. Light corpuscles, going near to
a sharp edge, experienced, according to their ideas,
instantaneous (in today's terminology {\it nonlocal})
interactions with the edge, and that caused a
phenomenon they called {\it inflexion}. Today that phenomenon
is called diffraction, and the name has changed because Young,
Fresnel, Faraday and Maxwell taught us that nonlocal
``explanations'' are not explanations at all. What they
gave us instead was a consistently {\it wave} explanation
of diffraction and interference, and theirs remains the only
explanation of those phenomena right up to the present day.

So how does it come to pass that the strongest claims
to have observed ``quantum nonlocality'' now come from
certain opticians? We suggest that it is because certain
opticians have allowed themselves to be seduced by
formal elegance, just like Newton's immediate successors.
Indeed, just like their intellectual ancestors, they have
been carried away by a formally elegant {\it mechanical}
theory. Yes, quantum {\it mechanics} is elegant, but only
as long as it applies to systems with a few degrees of
freedom. Light fields have infinite degrees of freedom,
and a mature treatment of them requires the considerably less
elegant apparatus of {\it quantum field theory} - not only
less elegant, but bristling with all sorts of problems
associated with divergences and renormalizations.

Will it be necessary to abandon the quantum formalism in
order to obtain a local description of ``multiphoton''
processes? We cannot yet give a complete answer to this
question, but we do assert that, with a certain natural
extension of the term ``classical'', all of the light
fields, including those currently classified as
``nonclassical'', which have so far been produced
in the laboratory are, in fact, entirely classical;
they are adequately described by the {\it unquantized}
Maxwell equations. We shall see that, in order to
extend our notion of ``classical'', it will be
necessary first to escape from Hilbert space and
place ourselves in classical phase space; this
enables us to adopt the point of view, which has
been anathematized by the Copenhagen school, that
electromagnetic waves are real waves, in ordinary space
and time, having both amplitude and phase. Whereas it
may seem natural, as long as we are imprisoned in Hilbert
space, to think of ``photons'', created at one point
and absorbed at another, the phase-space description
we advocate keeps us entirely within the confines
of classical electromagnetic theory. We believe that
this step has already been taken, but not fully
acknowledged, by a substantial part of the
quantum-optics community. For example, three
review articles $^{(2-4)}$ on light
squeezing make extensive use of phase-space
diagrams, and  one of them$^{(3)}$ states explicitly that the photon
description of the light field is not helpful in
the understanding of the phenomenon. We now propose
to extend this judgement to the light emitted in
atomic-cascade and parametric-down-conversion
processes, as well as the microwave radiation
contained in cavities.

\section{What is a classical light field?}

At the moment the accepted convention is to define as
``classical'', or more precisely ``Glauber-classical'',
a field which is a mixture of pure coherent states. For
a single mode of the field, the density matrix of
such a state is $^{(5)}$
\eqb
\hat\rho=\int\mid\alpha\rangle P(\alpha)\langle\alpha\mid
d^2\alpha,
\end{equation}
where $P$ is nonnegative and
\begin{equation}
\mid\alpha\rangle=\exp(\alpha\hat a^+-\alpha^*\hat a)
\mid 0\rangle.
\end{equation}
As we have argued elsewhere $^{(6-9)}$, there is a strong
case for extending the notion of ``classical'' to the set
of states whose Wigner density is nonnegative. All
Glauber-classical states are classical in this wider
sense. Indeed the Wigner density of the above single-mode
state is
\begin{equation}
W(\alpha)=2\pi^{-1}\int P(\alpha')\exp(-2\mid \alpha-\alpha'\mid^2)
d^2\alpha'.
\end{equation}

Any quantum state, defined by a density matrix, has a
well defined Wigner density, but in general there is no
guarantee that this density will be a nonnegative function.
The vacuum state is Glauber-classical with
\begin{equation}
P(\alpha)=\delta^2(\alpha),
\end{equation}
and therefore with
\begin{equation}
W(\alpha)=2\pi^{-1}\exp(-2\mid \alpha \mid ^2).
\end{equation}

According to some experts$^{(10,11)}$, the difference
between the $P$- and $W$-representations is entirely formal,
but we disagree$^{(9)}$. Equation (4) suggests that
the vacuum is truly empty, whereas equation (5)
suggests that there is a real nontrivial
distribution of mode amplitudes and phases in the
vacuum. Indeed such a view suggests that the {\it word}
``vacuum'' is (like ``inflexion'') obsolete, and that
we should call it something else (like ``plenum'').
This point of view is implicitly supported in that
part of the quantum-optics community which takes
phase-space diagrams seriously$^{(2-4)}$. More
explicitly, the concept of the real zeropoint field
has been central to {\it stochastic electrodynamics}
since the early 1960s$^{(12-14)}$, and its role
has been acknowledged more recently in quantum
electrodynamics$^{(15)}$.

We consider Max Planck$^{(16)}$ to be
the originator, in 1911, of the real zeropoint field, so we shall
call light fields with nonnegative Wigner densities
``Planck-classical'', and will consider a field to
be nonclassical only if it is not Planck-classical.
There are some Planck-classical fields
which are not Glauber-classical. An outstanding example
of such a field is the
squeezed vacuum state
\begin{equation}
\mid \zeta\rangle = exp(\zeta \hat a^{+^2}-\zeta^*\hat a^2)
\mid 0\rangle.
\end{equation}

So, using our new definition, are there {\it any} nonclassical states
of the light field? The simple answer is, of course, Yes. Indeed
most states of the Hilbert space are nonclassical,
 because, from a theorem
of Hudson$^{(17)}$, generalized by Soto and Claverie$^{(18)}$ ,
no pure states other than the Gaussian subset have
nonnegative Wigner density. In particular, the single-mode
one-photon Fock state
\begin{equation}
\mid 1\rangle = \hat a^+\mid0\rangle
\end{equation}
has Wigner density
\begin{equation}
W_1(\alpha)=2\pi^{-1}(4\mid\alpha\mid^2-1)
\exp(-2\mid\alpha\mid^2).
\end{equation}

With respect to the ``nonclassical'' states of the
light field currently widely reported as having been
observed, our response is that something approximating
the squeezed vacuum, as described by equation (6),
{\it has} been observed; this, however, according to
our new classification, is a {\it classical} state,
though not Glauber-classical. As for Fock states,
represented, for example, by equation (7), we consider
that the claims to have observed them are incorrect,
and that discussions on such exotic properties
(quantum nonlocality, entanglement etc.), which
such states would have, if they were to exist,
are misguided.

\section{Is the ``one-photon'' state classical?}
We have just seen that the {\it single-mode}
one-photon state, represented by equation (7),
is not Planck-classical. But we find it amazing that
anyone$^{(19)}$ should try to discuss such questions as
locality and causality on the basis of waves
which fill the whole of space and time! The
single-mode representation of a real atomic
signal is clearly inadequate.

If we wish to represent the output of a {\it real}
atomic source, we must take account not only of the
fact that each atomic light signal occupies a finite
time interval (typically about 5ns), but also that
neither the time nor the direction of the emitted
radiation can be controlled. (We are advocating a
return to an unambiguously {\it wave} description of
light, so any signal is emitted into a range of directions.
Nevertheless, the spatial distribution of the signal
will be influenced, for example, by the atom's charge
distribution at the time of emission; this cannot be
controlled.)

The first of these requirements leads us to a multimode
description of the light signal, while the second forces
us to abandon its description as a pure state (that is a
vector in Hilbert space) and use, instead, a density matrix.
We have shown$^{(20)}$ that, after incorporating these
two features, the density matrix is that of a chaotic state,
that is its Wigner density is Gaussian
and the state is classical. If, however,
the signal is part of an atomic-cascade process,
it is possible to use one signal in the cascade to
monitor the observation time of its partner, as in the
experiment of Grangier, Roger and Aspect$^{(21)}$.
In that case$^{(20)}$ we must use a different Wigner
density - again positive, however -
for the subensemble of monitored signals, and
this, as we shall show in a later section, allows a
purely wave explanation of what
Grangier, Roger and Aspect thought was
corpuscular behaviour.
\section{The role of the zeropoint field}
The real zeropoint field plays a crucial role in
explaining how purely wave phenomena may be
misinterpreted as evidence of corpuscular
behaviour. Recognition of its role would be
a convincing vindication of Max Planck$^{(16)}$, because
he introduced the concept of the zeropoint field
precisely in order to oppose Einstein's {\it Lichtquanten},
which were the forerunners of photons.

We consider first the way in which a theory with a
real zeropoint field views the action of a beam splitter.
Such a device was used by Grangier, Roger and Aspect$^{(21)}$
to demonstrate a phenomenon they called {\it anticorrelation}
in the outgoing channels. If we consider the ``vacuum''
to be empty, then it seems almost unavoidable to
assume that the intensity of any incoming classical
signal is equal to the sum of the intensities in the
outgoing channels, and also that the detection
probability in both channels is proportional to the
signal intensities in those channels. With such a
description it is not possible to explain the
anticorrelation data; these were interpreted by
Grangier, Roger and Aspect as evidence that the
whole ``photon'' goes into one or other of the outgoing
channels. It is easy to see, qualitatively, how the
explanatory power of a purely wave theory is increased by
the recognition of a real zeropoint field. A beam splitter
does not simply split an incoming wave into two parts;
it {\it mixes} together {\it two} incoming waves, one of
them from the ``vacuum'', to give the two outgoing waves
(see Figure 1).

\begin{center}

\unitlength=1mm
\begin{picture}(60,40)

\put(0,20){\line(1,0){60}}
\put(30,20){\line(0,-1){20}}
\multiput(30,20)(0,4){5}{\line(0,1){2}}
\put(15,20){\vector(1,0){0}}
\put(45,20){\vector(1,0){0}}
\put(30,10){\vector(0,-1){0}}
\put(30,32){\vector(0,-1){0}}
\put(18,17){\makebox(0,0){$a$}}
\put(48,17){\makebox(0,0){$b$}}
\put(33,29){\makebox(0,0){$d$}}
\put(33,7){\makebox(0,0){$c$}}
\thicklines
\put(25,25){\line(1,-1){10}}
\end{picture}

\end{center}

\begin{list}{}{\leftmargin=3em \rightmargin=3em}\item[]
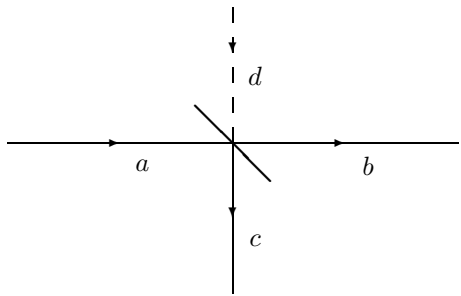
{\bf Figure 1}: The beam splitter mixes the incoming
signal ($a$) with the relevant modes of the zeropoint field ($d$)
to give the signals ($b$ and $c$) in the two outgoing
channels.
\end{list}

Something similar occurs in a nonlinear optical crystal.
An intense coherent input causes two modes of the
zeropoint field, initially uncorrelated, to become
both enhanced in their amplitudes and correlated
(see Figure 2). This
in turn causes correlated
``photon''  counts in the outgoing channels. The current
name for what occurs in the crystal is {\it
parametric down conversion} but this is yet another
example (like ``inflexion") of a bad concept - the
``photon''  - giving rise to a misleading name and
description; it describes an incoming photon of the
coherent beam as converting into two completely
new photons. But all modes of the field are
{\it already present} before the intervention of
the coherent beam and the nonlinear crystal. The
{\it down conversion} is, more accurately, a
{\it correlated amplification} of certain modes
of the zeropoint field.

\begin{center}

\unitlength=1mm
\begin{picture}(90,60)

\put(45,32.5){\makebox(0,0){{\footnotesize nonlinear}}}
\put(45,27.5){\makebox(0,0){{\footnotesize crystal}}}
\put(52.5,35){\line(3,2){37.5}}
\put(52.5,25){\line(3,-2){37.5}}
\multiput(37.5,25)(-9,-6){4}{\line(-3,-2){4.5}}
\multiput(37.5,35)(-9,6){4}{\line(-3,2){4.5}}
\put(20,30){\vector(1,0){0}}
\put(21,46){\vector(3,-2){0}}
\put(21,14){\vector(3,2){0}}
\put(69,46){\vector(3,2){0}}
\put(69,14){\vector(3,-2){0}}
\put(20,26){\makebox(0,0){coherent beam}}
\put(25,14){\makebox(0,0){$E_0'$}}
\put(25,46){\makebox(0,0){$E_0$}}
\put(60,14){\makebox(0,0){signal}}
\put(60,46){\makebox(0,0){idler}}
\thicklines
\put(0,30){\line(1,0){37.5}}
\put(37.5,25){\line(1,0){15}}
\put(37.5,35){\line(1,0){15}}
\put(37.5,25){\line(0,1){10}}
\put(52.5,25){\line(0,1){10}}
\end{picture}
\end{center}

\begin{list}{}{\leftmargin=3em \rightmargin=3em}\item[]
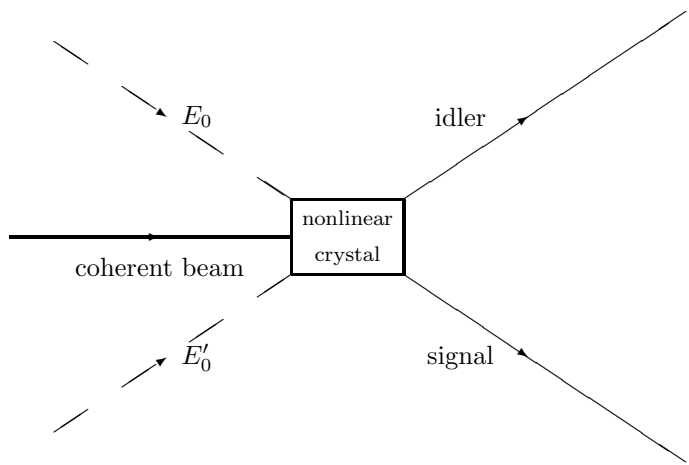
{\bf Figure 2}: The coherent beam modifies the two
initially uncorrelated zeropoint amplitudes $E_0$ and
$E_0'$ to produce the two correlated (``signal''  and ``idler'')
outputs.
\end{list}

The two outgoing signals from the beam splitter in Figure 1
are given by$^{(22)}$
\begin{eqnarray}
E_b&=&TE_a+iRE_d,\\
E_c&=&TE_d+iRE_a,
\end{eqnarray}
where $R$ and $T$ are real coefficients satisfying
\begin{equation}
R^2+T^2=1.
\end{equation}
The two outgoing signals from the nonlinear crystal in Figure 2
are given by$^{(22)}$
\begin{eqnarray}
E_{{\rm signal}}&=&E_0+gVE_0^{'*},\\
E_{{\rm idler}}&=&E'_0+gVE_0^*,
\end{eqnarray}
where $g$ is a coupling constant and $V$ is the analytic signal
of the coherent beam. Because these outputs are linearly related to
the inputs, we have been able to deduce that the joint Wigner
density of the outgoing beams is positive, both when the inputs
are all Gaussian, as in Figure 2, and when one of the inputs is
``single-photon'', as would be the case in the Grangier-Roger-Aspect
experiment. We remark that the linearity property holds always in
the beam-splitter case, but that it holds only to first-order
perturbation approximation in the nonlinear-crystal case. In the
full quantum formalism, higher-order effects
\footnote{More accurately, the linearity property holds
in the approximation where we may neglect changes in
the laser intensity - so-called {\it depletion} -
resulting from absorption in the crystal. Within
this approximation, the linearity property holds
to all orders of perturbation theory. {\it This
footnote was added after submittal of the article
to the editors of the Vigier Conference
Proceedings.}}
could, possibly,
give an outgoing Wigner density taking negative values$^{(9,10)}$,
but such effects are, at present, not experimentally
observable.

\section{The theory of detection}
We have just seen that taking account of the zeropoint
field leads us to a different understanding of certain optical
devices. In particular, the recognition of the previously
``missing'' inputs, as in Figure 1 and Figure 2, means that
the sum of the intensities of the outgoing signals is not
equal to the intensity of just one incoming signal. This
new feature of a zeropoint field theory is sufficient to take
away the mystery of {\it enhancement} at a beam splitter.
Applying this idea to a {\it polarizing} beam splitter,
we have been able to show$^{(23,24)}$ that all the ``nonlocal''
data for polarizations of light signals from atomic
cascades$^{(25)}$ have a local explanation.
However, it is now necessary to modify the theory of detection.
All previous semiclassical theories$^{(26)}$ have omitted the
zeropoint field, and it has been assumed that the detection
probability is proportional to the signal intensity. Since
the total intensity in all the zeropoint modes is enormously
greater than any signals, it must follow that all detectors
are ``blind'', or nearly so, to the zeropoint intensity. This
must be so even when the ``signal'' is the light from the
Sun and the detectors are our own eyes!

The subtraction of the zeropoint noise is, we claim,
already a feature of the standard theory, in which
{\it light detectors are considered to be
normal-ordering devices$^{(10)}$}. The probability
of joint detection in the two outgoing channels of
Figure 1 is given by$^{(5)}$
\begin{equation}
{\rm Pr[\,joint\, detection\,]}=\eta_b\eta_c\int_0^\tau dt
\int_0^\tau dt'
\langle{\rm N}[\hat I_b(t)\hat I_c(t')]\rangle,
\end{equation}
where N denotes that the field amplitudes in
$\hat I_b$ and $\hat I_c$ are normally ordered,
and $\eta_b,\,\eta_c$ are the detector efficiencies.
We have shown that an equivalent expression is
\begin{equation}
{\rm Pr[\,joint\,detection\,]}=\eta_b\eta_c\int_0^\tau dt
\int_0^\tau dt'
\langle{\rm S}[\{\hat I_b(t)-I_0\}\{\hat I_c(t')-I_0\}]
\rangle ,
\end{equation}
where S denotes a symmetric ordering of the field
amplitudes, and $I_0$ is the zeropoint intensity in the
relevant modes. This enabled us to replace the whole
expression by an integral$^{(5)}$, over the classical phase
space, involving the (positive) Wigner density, and hence
give a purely wave explanation of the anticorrelation of
the photoelectron counts in these channels. We have made
a similar analysis$^{(22)}$ of the correlated signals in
Figure 2, and hence have been able to give a purely
wave explanation of the experiments (some of which
the authors describe as ``mind boggling'') described
in Reference$^{(19)}$.

It remains a problem to explain how this formal
subtraction of the zeropoint is actually achieved by the
detectors. There must exist some positive functional of
the field amplitudes whose average value, weighted
by the Wigner density, is very small when only the
zeropoint is present. Note that real detectors all have
a finite dark rate, so the zeropoint will always
give {\it some} detection events; hence the theory
of detection we are demanding will actually
explain more than the current theory.

In the absence of such a theory we constructed
a simpler model theory$^{(23,24)}$ in order to
illustrate how the noise subtraction, in
combination with the enhancement mechanism
described in the previous section, gave rise
to certain ``particle-like'' counting statistics
in the two channels.

The explanation of how detectors are able to extract
signals from the very large zeropoint background is
a very difficult problem which we have not yet managed to
solve. The day that theoretical physicists begin
seriously to confront it is when they will begin, perhaps,
to recover the respect of the rest of the scientific
community. Despite our failure to resolve it, we state
our conclusions, namely that the photon is obsolete,
that light is nothing but waves, and that all wave
fields fluctuate (see the opening sentence of
Reference$^{(4)}$). The next section is nothing but
a postscript to this conclusion.

\section{The microwave field in a cavity}
Proposals have been made for the construction of
experimental situations resembling the Schroedinger
cat$^{(27,28)}$, in which two quite large objects,
namely a Rydberg atom and a microwave cavity, are
put in a superposition state, so that certain of
their properties are ``entangled''.

Since it is not, at present, possible to observe
directly the state of the cavity, this entanglement
(if it really existed!) could not be demonstrated
so readily as would be the case, with ``perfect''
detectors, for entangled light signals. Hence, any
serious attempt to construct a Schroedinger cat
must either seek to entangle two successive
Rydberg atoms passing through the same cavity$^{(29)}$
or make some plausible additional assumption about
the single Rydberg atom. We have taken the latter
course$^{(30)}$, since we are sure that it leads to
experimental requirements which are easily achievable
with current technology. The additional hypothesis
we propose is that the two-state stochastic process
represented by the Rydberg atom be {\it stationary}.
With this condition the Rydberg atom may be treated as
a macroscopic system - it is bigger than a protein molecule -
and the inequalities for such systems, deduced by
Leggett and Garg$^{(31)}$ (also called temporal
Bell inequalities) should apply. With the very
high Q-values claimed by experimenters
for the cavity, it should be possible, according to
current theory, to demonstrate a violation of the
Leggett-Garg inequality, but our analysis of the
data$^{(32)}$ so far available shows no such violation.
In the light of our experience with atomic cascades,
one should be modest about the conclusions one
draws. Hypotheses which seem plausible before
doing an experiment should, properly, often be
rejected in the light of the new evidence. This was
our experience with Clauser and Horne's$^{(33)}$
hypothesis of no enhancement at a beam splitter.
For the moment our inclination is to persist with
the stationarity hypothesis for the Rydberg atom.
It seems to us highly probable that the Q-values
currently claimed for supercooled cavities may
not take fully into account all the possible
relaxation mechanisms for the radiation in the
cavity, and it would not take much relaxation
to convert the ``ideal'' quantum electrodynamic
autocorrelation of the process into one satisfying
the Leggett-Garg inequalities.

We wish Jean-Pierre Vigier a very  happy birthday.\vspace {1cm}\\

\noindent
{\bf Acknowledgement}\\
We acknowledge financial assistance of DGICYT Project No. PB-92-0507
(Spain).\vspace{1cm}\\
\noindent
{\bf References}\\
1. T. W. Marshall, Found. Phys. {\bf 22}, 363 (1992).\\
2. R. E. Slusher and B. Yurke, Sci. Amer. {\bf 258}, 32 (1988).\\
3. E. Giacobino, C. Fabre, A. Heidmann and S. Reynaud,
La Recherche {\bf 21}, 170 (1990).\\
4. R. Loudon and P. L. Knight, J. Mod. Opt. {\bf 34}, 709 (1987).\\
5. J. Pe\v rina, {\it Quantum Statistics of Linear and
Nonlinear Optical Phenomena} (Reidel, Dordrecht,1984).\\
6. T. W. Marshall and E. Santos, Found. Phys. Lett. {\bf 5} ,573
(1992).\\
7. T. W. Marshall, E. Santos and A. Vidiella-Barranco,
{\it Proc. 3rd Int. Workshop on
Squeezed States and Uncertainty Relations} (D. Han, Y. S. Kim,
N. H. Rubin, Y. Shih, W. W. Zachary  eds.) (NASA Conf. Series, No. 3270, 1994) page 581.\\
8. T. W. Marshall and E. Santos, Phys. Rev. A {\bf 41}, 1582
(1990).\\
9. T. W. Marshall, Phys. Rev. A {\bf 44}, 7854 (1991).\\
10.P. Kinsler and P. D. Drummond, Phys. Rev. A {\bf 44}, 7848
(1991). \\
11.P. Milonni, {\it The Quantum Vacuum} (Academic, San Diego,1993)
page 142.\\
12.T. H. Boyer, in {\it Foundations of Radiation Theory
and Quantum Electrodynamics}\\(A. O. Barut, ed.)
(Plenum, New York, 1980).\\
13.L. de la Pe\~na, in {\it Stochastic Processes Applied
to Physics and Other Related Fields}\\(B. Gomez, S. M. Moore,
A. M. Rodriguez-Vargas and A. Rueda, eds.)
(World Scientific, Singapore, 1983). \\
14.T. H. Boyer, Sci. Amer. August 1985. \\
15.J. Dalibard, J. Dupont-Roc and C. Cohen-Tannoudji,
J. Phys. (Paris) {\bf 43}, 1617 (1982).\\
16.M. Planck, {\it Theory of Heat Radiation}
(Dover, New York, 1959).\\
17.R. L. Hudson, Rep. Math. Phys. {\bf 6}, 249 (1974).\\
18.F. Soto and P. Claverie, J. Math. Phys. {\bf 24}, 97 (1983). \\
19.D. M. Greenberger, M. A. Horne and A.  Zeilinger,
Phys. Today {\bf 46}, 22 (1993). \\
20.T. W. Marshall, in {\it Fundamental Problems in Quantum Physics}\\
(M. Ferrero and A. van der Merwe, eds.) (Kluwer, Dordrecht, 1995) page 187.\\
21.P. Grangier, G. Roger and A. Aspect, Europhys. Lett.
{\bf 1}, 173 (1986).\\
22.A. Casado, T. W. Marshall and E. Santos, preprint Univ. de Cantabria\\
FMESC 3 (1995).\\
23.T. W. Marshall and E. Santos, Found. Phys. {\bf 18}, 185 (1988).\\
24.T. W. Marshall and E. Santos, Phys. Rev. A {\bf 39}, 6271 (1989).\\
25.A. J. Duncan and H. Kleinpoppen, in {\it Quantum
Mechanics versus Local Realism}\\ (F. Selleri, ed.)
(Plenum, New York, 1988).\\
26.L. Mandel, Prog. in Optics {\bf 13}, 27 (1976).\\
27.E. Schroedinger, Proc. Camb. Phil. Soc. {\bf 31}, 555 (1935).\\
28.S. Haroche, M. Brune, J. M. Raimond and L. Davidovich,
in {\it Fundamentals in Quantum Optics} (F. Ehlotzki, ed.)
(Springer, Berlin, 1993).\\
29.S. J. D. Phoenix and S. M. Barnett, J. Mod. Opt.
{\bf 40}, 979 (1993).\\
30.S. F. Huelga, T. W. Marshall and E. Santos,
Phys. Rev. A {\bf 52}, R2497-R2500 (1995).\\
31.A. J. Leggett and A. Garg, Phys. Rev. Lett. {\bf 54}, 587 (1985).
\\
32.G. Rempe, H. Walther and N. Klein, Phys. Rev. Lett.
{\bf 58}, 353 (1987).\\
33.J. F. Clauser and M. A. Horne, Phys. Rev. D {\bf 10}, 526 (1974).\\
\end{document}